\documentclass[eqsecnum,floats,aps,nofootinbib]{revtex4}
\usepackage{amsmath,amssymb,latexsym,epsfig}

\newcommand{\speq}{\!\!& = &\!\!}
\def\nn{\nonumber}

\begin{document}
\thispagestyle{empty}

\title{A note on the extension of the Dirac method}

\author{D. Babusci}
\email{danilo.babusci@lnf.infn.it}
\affiliation{INFN - Laboratori Nazionali di Frascati, v.le E. Fermi, 40, 00044 
Frascati (Roma), Italy} 
\author{G. Dattoli}
\email{giuseppe.dattoli@enea.it}
\author{M. Quattromini}
\email{quattromini@frascati.enea.it}
\affiliation{ENEA -  Centro Ricerche Frascati, v.le E. Fermi, 45, 00044,  
Frascati (Roma), Italy}
\author{P. E. Ricci}
\email{paoloemilioricci@gmail.com}
\affiliation{International Telematic University UniNettuno, 
Corso Vittorio Emanuele II, 39, 00186 Roma, Italy}

\begin{abstract}
In this note we extend the Dirac method to partial differential equations involving higher order roots of differential operators.
\end{abstract}

\maketitle

\section{Introduction} \label{sec:intro}
In deriving his celebrated equation, the Dirac stroke of genius was to use the anti-commuting nature of the Pauli matrices 
\begin{equation}
\left\{\hat{\sigma}_j,\hat{\sigma}_k\right\} \,=\, 2\,\delta_{jk}\,, \qquad\qquad (j, k = 1, 2, 3)
\end{equation}
to rewrite the Pythagorean theorem as follows
\begin{equation}
\label{pytha}
A = \sqrt{B^2 + C^2} = B\,\hat{\sigma}_j + C\,\hat{\sigma}_k \qquad\qquad (j \neq k)
\end{equation}
where $A, B, C$ can be either numbers or operators. Eq.\ref{pytha} allows to eliminate the square root from the relativistic Schr\"odinger 
equation
\begin{equation}
\imath\,\hbar\,\partial_t\,\Psi = c\,\sqrt{\hat{p}^2 + m^2\,c^2}\,\Psi\,,
\end{equation}
which, can accordingly be written as
\begin{equation}
\imath\,\hbar\,\partial_t\,\underline{\Psi} = c\,(\hat{p}\,\hat{\sigma}_1 + m\,c\,\hat{\sigma}_3)\,\underline{\Psi}\,.
\end{equation}
where $\underline{\Psi}$ is a 2-component vector. This equation resembles the Dirac equation, but it should not be confused with it 
or with the Pauli equation. It does not involve, indeed, 4-component spinors and $4 \times 4$ matrices, and is interpreted as the equation 
ruling the free evolution of a relativistic spinless particle. The underlying physical problems have already been discussed in Refs. \cite{Wang,Bab1}, 
and will not be reconsidered here. 

What we are concerned is whether it is possible to generalize eq. (\ref{pytha}) as follows
\begin{equation}
A = \sqrt[3]{B^3 + C^3} = B\,\hat{\tau}_j + C\,\hat{\tau}_k \qquad\qquad (j \neq k)
\end{equation}
with the matrices $\hat{\tau}_j$ $(j = 1, 2, 3)$ satisfying the conditions
\begin{equation}
\label{taum}
\hat{\tau}_j^3 = \hat{1} \qquad\qquad \hat{\tau}_j^2\,\hat{\tau}_k + \hat{\tau}_j\,\hat{\tau}_k\,\hat{\tau}_j +
\hat{\tau}_k\,\hat{\tau}_j^2 = 0\,.
\end{equation}
If do exist, these matrices have to be associated with the cubic roots of unity
\begin{equation}
\epsilon_0 = 1 \qquad\qquad \epsilon_\pm = - \frac12 \pm \imath\,\frac{\sqrt{3}}2
\end{equation}
that satisfy the conditions
\begin{equation}
\epsilon^3 = 1\,, \qquad \epsilon_0 + \epsilon_+ + \epsilon_- = 0\,, \qquad \epsilon_+\,\epsilon_- = 1\,.
\end{equation}
We construct two of the possible matrices as
\begin{equation}
\label{mat12}
\hat{\tau}_1 = \left(
\begin{array}{ccc}
   0  &  1  &  0  \\
   0  &  0  &  1  \\
   1  &  0  &  0
\end{array}
\right)\,, \qquad\qquad 
\hat{\tau}_2 = \left(
\begin{array}{ccc}
   0  &  \epsilon_+  &  0  \\
   0  &  0  &  \epsilon_-  \\
   1  &  0  &  0
\end{array}
\right)\,, 
\end{equation}
which are easily shown to satisfy the conditions (\ref{taum}). Accordingly the equation
\begin{equation}
\partial_t\,\Phi = \sqrt[3]{\partial_x^3 + k}\,\Phi
\end{equation}
with $k$ being a constant, can be cast in the form
\begin{equation}
\label{evol}
\partial_t\,\underline{\Phi} = (\partial_x\,\hat{\tau}_1 + k\,\hat{\tau}_2)\,\underline{\Phi} \qquad\qquad 
\underline{\Phi} (t = 0) = \underline{\Phi}_0\,,
\end{equation}
where $\underline{\Phi}$ is a 3-component vector. This increase of dimensionality is the price to be paid if we want to use a Dirac-like 
procedure to avoid the use of operators raised to a non integer number (further comments on this point can be found in Ref. \cite{Bab1}).

The solution of eq. (\ref{evol}) can be written as
\begin{equation}
\label{phis}
\underline{\Phi} = \exp\left\{t\,(\partial_x\,\hat{\tau}_1 + k\,\hat{\tau}_2)\right\}\,\underline{\Phi}_0\,.
\end{equation}
The exponential operator in the last equation cannot be naively disentangled because the matrices $\hat{\tau}$ are not commuting. Indeed, one has
\begin{equation}
[\hat{\tau}_1, \hat{\tau}_2] = - \imath\,\sqrt{3}\,\hat{\tau}_3 
\end{equation}
with
\begin{equation}
\label{mat3}
\hat{\tau}_3 = \left(
\begin{array}{ccc}
   0  &  0  &  1  \\
   \epsilon_+   &  0  &  0 \\ 
   0  &  \epsilon_-  &  0  
\end{array}
\right)
\end{equation}
Also this matrix satisfies the conditions (\ref{taum}). 

The explicit solution of our problem can be found using operator ordering methods \cite{Dat1}. However, just to provide a feeling on the possibilities offered 
by the present method, we use the Trotter formula \cite{Trot}
\begin{equation}
\exp\left(\hat{A} + \hat{B}\right) = \lim_{n \to \infty} \left[\exp\left(\frac{\hat{A}}n\right)\,\exp\left(\frac{\hat{B}}n\right)\right]^n
\end{equation}
to write eq. (\ref{phis}) as follows
\begin{equation}
\underline{\Phi} = \lim_{n \to \infty} \left[\exp\left(\frac{t\,k}n\,\hat{\tau_2}\right)\,\exp\left(\frac{t\,\partial_x}n\,\hat{\tau_1}\right)\right]^n\,\underline{\Phi}_0\,.
\end{equation}
By considering small time intervals $\delta = t/n$, and evaluating the effect of the exponentials step by step, we get an iterative solution in the form
\begin{equation}
\underline{\Phi}_n = \exp\left(\delta\,k\,\hat{\tau_2}\right)\,\exp\left(\delta\,\partial_x\,\hat{\tau_1}\right)\,\underline{\Phi}_{n - 1}\,.
\end{equation}
The action of the evolution operator on the function on its right can be easily obtained by noting that the exponentials can be expanded in finite terms, as a 
consequence of the cyclical properties of the matrices $\hat{\tau}$, namely $\hat{\tau}^{3 m} = 1$, thus getting \cite{Dat2}
\begin{equation}
\label{ }
\exp\left(\alpha\,\hat{\tau}_j\right) = C_0 (\alpha)\,\hat{1} + C_1 (\alpha)\,\hat{\tau}_j + C_2 (\alpha)\,\hat{\tau}_j^2\,,
\end{equation}
with
\begin{equation}
\label{cmal}
C_m (\alpha) = \sum_{l = 0}^\infty \frac{\alpha^{\,3\,l + m}}{(3\,l + m)!}\,\qquad\qquad (m = 0, 1, 2)\,.
\end{equation}
These functions are recognized as the pseudo-hyperbolic functions introduced in \cite{Ricc}, or, in more general terms, as the generalized trigonometric functions 
discussed recently in Refs. \cite{Yama,Bab2}.

According to the previous relations, at the first iteration we get ($\underline{\Phi}_0 = (\varphi_1 (x), \varphi_2 (x), \varphi_3 (x))^T$) 
\begin{eqnarray}
\underline{\Phi}_1 \speq \exp\left(\delta\,k\,\hat{\tau_2}\right)\,\exp\left(\delta\,\partial_x\,\hat{\tau_1}\right)\,\underline{\Phi}_0 \nn \\
\speq \left(
          \begin{array}{ccc}
          C_0 (\delta\,k)  &  \epsilon_+\,C_1 (\delta\,k)  &  C_2 (\delta\,k) \\
          \epsilon_-\,C_2 (\delta\,k)  & C_0 (\delta\,k)  & \epsilon_-\,C_1 (\delta\,k)  \\
          C_1 (\delta\,k)  &  \epsilon_+\,C_2 (\delta\,k)  & C_0 (\delta\,k)
          \end{array}
          \right)\,
          \left( 
          \begin{array}{c}
          C_0 (\delta\,\partial_x)\,\varphi_1 (x) + C_1 (\delta\,\partial_x)\,\varphi_2 (x) + C_2 (\delta\,\partial_x)\,\varphi_3 (x) \\
          C_2 (\delta\,\partial_x)\,\varphi_1 (x) + C_0 (\delta\,\partial_x)\,\varphi_2 (x) + C_1 (\delta\,\partial_x)\,\varphi_3 (x) \\
          C_1 (\delta\,\partial_x)\,\varphi_1 (x) + C_2 (\delta\,\partial_x)\,\varphi_2 (x) + C_0 (\delta\,\partial_x)\,\varphi_3 (x) 
          \end{array}
          \right)\,.
\end{eqnarray}
The successive iterations are straightforwardly obtained along the same lines.

It is evident that the method we have outlined open new perspectives and is amenable for further implementations which will carefully be discussed in a forthcoming 
investigation where the case of higher order roots of unity will be discussed.




\end{document}